%% file: paper.tex
\documentclass[conference]{IEEEtran}

\usepackage{graphicx}
\usepackage{amsmath}

\IEEEoverridecommandlockouts
\begin{document}

\title{DeepRS: Deep-learning Based Network-Adaptive FEC for Real-Time Video Communications}

\author{
Sheng Cheng, Han Hu, Xinggong Zhang\textsuperscript{*}, Zongming Guo\\
Wangxuan Institute of Computer Technology\\ 
Peking University\\
Email: \{chaser\_wind, huhan951753, zhangxg, guozongming\}@pku.edu.cn%
\thanks{\textsuperscript{*}Dr. Xinggong Zhang is the corresponding author.}%
\thanks{This work was supported by the National Key R\&D program of China (2019YFB1802701, 2018YFB0803702), Alibaba Innovative Research (AIR) Program (XT622018001708).}
}

\maketitle

\input{abstract}


%\IEEEpeerreviewmaketitle

\input{intro}
\input{system}

\input{eva}
\input{conclu}

\bibliography{paper}
\bibliographystyle{IEEEtran}

\end{document}

%% file: abstract.tex
\begin{abstract}
As real-time multimedia streaming thriving, Forward Error Correction (FEC) methods 
have been studied and applied extensively these years. Most of researchers paid their 
attention to the coding algorithms, attempted to balance the trade off between recovery 
ratio and delay with fewer redundance. However, when packet loss pattern 
changes dynamically, the redundance waste is too serious to be ignored. In this work, 
we propose a novel algorithm which adjusts the redundance ratio of FEC encoder 
according to the prediction of packet loss. Receivers are additionally required to feedback
observed packet loss pattern. Streaming sender collects the feedbacked
 packet loss pattern and predicts the number of packet loss in the incoming 
short period. As for implementation, we adopt long short-term memory (LSTM) network 
as our deep learning algorithm, and exquisitely embed it in our adaptive FEC system. 
With the extensive
experiments, our proposed scheme outperforms other FEC methods greatly both
in the simulations and evaluations on traces observed from the real world.
\end{abstract}

%% file: intro.tex
\section{Introduction}

Real-time video streaming is getting more and more prevalent in these years.
Network video streaming is expected to account for 82\% of Internet traffic by 2022. 
In the meanwhile, a growing share of network video will take the form of live streaming video \cite{vni}, 
and Real-time Video Communication (RTC) is drawing increasing attention of users and researchers. 
However, packet loss is proved to be a critical problem for RTC, 
because it could cause distortion and decoding error, thus degrading users'
Quality of Experience (QoE). 
% may be deleted
% As for the reliable transport protocols like Transmission Control Protocol (TCP), 
% there has been some wonderful coding schemes such as TCP/NC \cite{TCPNC} and CTCP \cite{CTCP}, 
% intended to achieve higher throughput and reduce packet loss. 
% On the contrast, as for User Datagram Protocol (UDP), the previous works do not perfectly solve the problems yet.
% may be deleted
Considering the demand of RTC applications like video conferences that one way delay is limited 
to no more than $200ms$ according to the standard of International Telecommunication Union (ITU), 
it is challenging to recover packet losses via restransmission on Internet.
Automatic Repeat-reQuest (ARQ) is a traditional error correction method, 
but its recovery delay is longer than an RTT.
%Due to strict requirement on recovery delay of real-time video streaming, 
%it is impractical to use conventional ARQ mechanism to recover lost packets in RTC.
In order to solve this problem efficiently, application-layer Forward Error Correction (FEC) 
\cite{wang1998error} has been proposed. 
Application-layer FEC applies error-resilient coding algorithms to source packets to generate redundant packets, 
which can recover the ever lost source packets with relatively low latency when packet loss happens.
It is validated that application-layer FEC is an effective technique to achieve
low-latency packet transmissioin, but it is still a big challenge for application-layer FEC 
to improve the efficiency under time-varying packet loss pattern. 
%Application-layer FEC is drawing more attention for its 
% With FEC enabled, the receiver
% of streaming is able to recover lost packets on the spot.
\begin{figure}[t]
	\centering
	\includegraphics[width=0.48\textwidth]{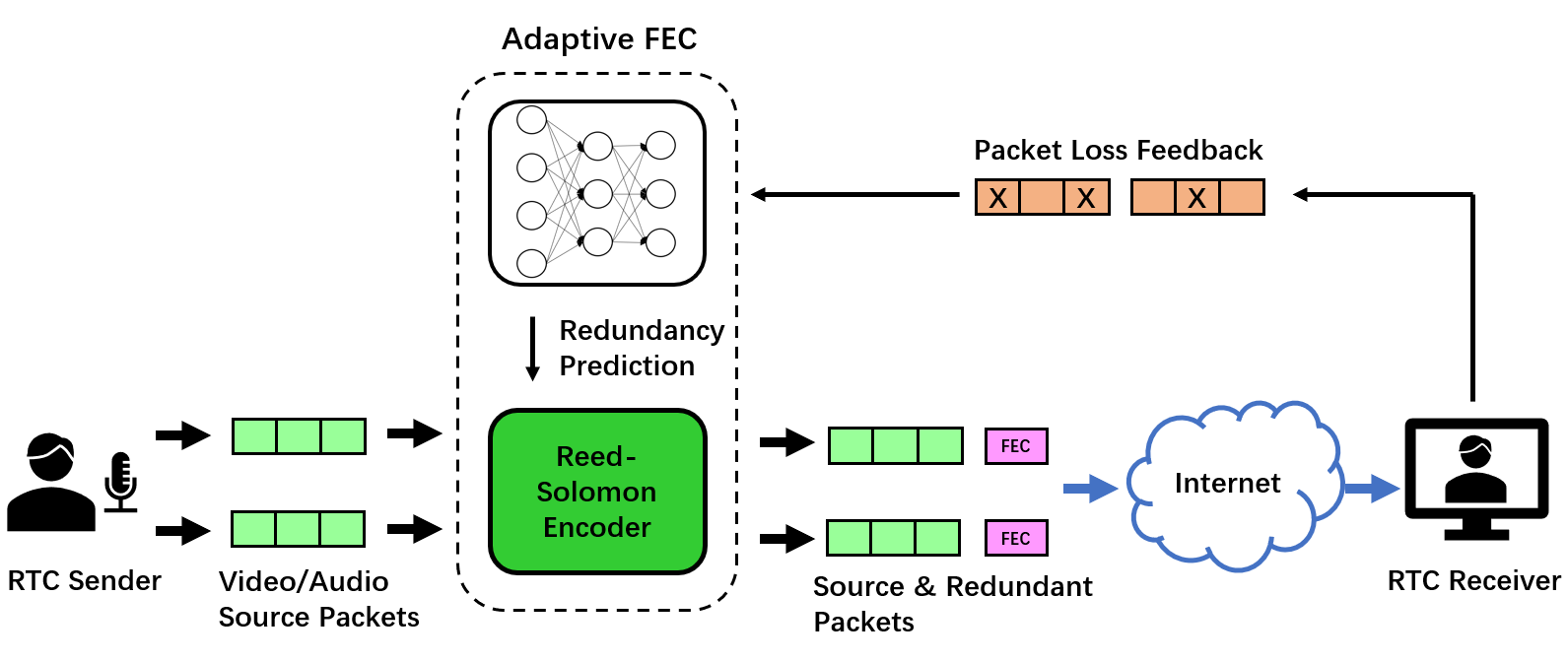}
	\vspace{0.8em}
	\caption{The framework of deep-learning based adaptive FEC. DeepRS collects delayed feedback from receiver, 
		makes prediction on future packet loss using LSTM model, and adjusts the parameters of RS encoder.}
	\label{hierachy}
	
\end{figure}

%FEC与coding的衔接。
There are two main categories of the existing works addressing the issue: coding scheme and adaptive FEC.
\begin{itemize}
    \item \textbf{Coding scheme} focuses on devising new FEC coding algorithms to enhance the
    efficiency of packet loss recovery. For instance, fountain code \cite{mackay2005fountain}, 
    Raptor code \cite{Demir2006Raptor, Bouras2012Evaluating} and Reed-Solomon (RS) code 
    \cite{wicker1999reed, sudan1997decoding} have been studied. Moreover,
    Xiao et.al. put forward the expanding window approach to apply unequal protection to 
    different packets of video/audio data on the basis of their importance, so as to improve QoE in RTC \cite{exp}.
    However, these works are based on the hypothesis that network packet loss pattern is regular or fixed, 
    which is not practical on real Internet. It is often the case that the redundant packets of some
    blocks are wasted due to the absence of packet loss, or they can not help owing to the overfull packet loss. 
    \item \textbf{Adaptive FEC} is first suggested to cope with the challenge of dynamic
    loss pattern by Padhye et.al \cite{ipadp}. Atiya et.al came up with a nonlinear prediction 
    approach to realize automatic feature selection \cite{Atiya2007Packet}, and Fong et.al. 
    combined an ingenious coding scheme along with a network adaptive algorithm for parameter update \cite{fong2019low}.
    Nevertheless, all of them simply take the historical network pattern as the prediction of future pattern, 
    ignoring the possible complicated relationship between history and future. When network
    conditions change frequently, these techniques will not work well as expected.
\end{itemize}

To make good use of the contextual relationship between history and future, 
we propose DeepRS, a novel FEC algorithm which predicts network packet loss with deep neural 
network, dynamically adjusts the redundancy ratio, 
and improves the efficiency of FEC scheme markedly. 
The framework of deep-learning based adaptive FEC is shown in Fig. \ref{hierachy}.
%In consideration of delay and effectiveness, we choose to use FEC scheme instead of ARQ. 
%[why choose RS code, how to realize network-adaptive]
DeepRS predicts the packet loss based on the feedback from receiver, 
determines the amount of redundant packets and applies RS coding algorithm to 
encode this video block.
In order to make the best of the contextual relevance of network loss pattern, we propose 
a prediction method of packet loss based on Long Short-Term Memory (LSTM) network.
A large number of experiments on simulation and real Internet traces show that 
the recovery ratio of DeepRS is $70\%$ higher than the compared algorithms in the case of a fixed
total redundancy rate, and DeepRS can realize adaptive FEC redundancy in any network dynamic. 
To the best of our knowledge, this is the first time to verify that 
the deep-learning FEC is able to improve FEC efficiency significantly. 
It may open a new door to error correction coding research.
% With the help of deep learning model, DeepRS is sensitive to changes of network conditions and
% performs well while predicting packet loss pattern on network.
% The LSTM module can effectively analyze the historical data and then predict the incoming loss pattern.
% RS code is selected as the basic coding algorithm for its robustness and 
% simplicity, especially its flexibility on redundancy adjustment.
% As a classical FEC coding algorithm, RS code satisfies our 
% need of capability to adjust redundancy ratio freely, and tends to be compatible with our LSTM model. 
% Besides, we decide to use packet-level RS code instead of symbol-level RS code, 
% since real network packets share different lengths which makes the latter uneasy to be applied.
% With high accuracy of prediction, DeepRS alters its redundancy ratio automatically to achieve more packet recovery
% with finite redundancy.

The main contributions of this paper are listed as follows.
\begin{enumerate}
    \item Design an LSTM model to solve the problem of predicting dynamic network packet loss.
    \item Come about the packet loss counting method to solve the problem of predicting loss pattern accurately.
    \item Propose the block gap prediction method to solve the problem of delayed feedback.
\end{enumerate}

The remainder of 
this paper is organized as follows. Section \ref{system} covers concrete
definition and construction of DeepRS. In Section \ref{eva}, we present 
the setup of our experiments as well as the evaluations. And eventually we come to the conclusion in Section \ref{conclu}.
%[The related works that could be cited: 2019MM \cite{fong2019low}, Yufeng geng \cite{geng2015unequal}, and the papers cited in \cite{geng2015unequal}.]

%% file: system.tex
\section{System Design}
\label{system}

\begin{figure}[t]
	\centering
	\includegraphics[width=0.45\textwidth]{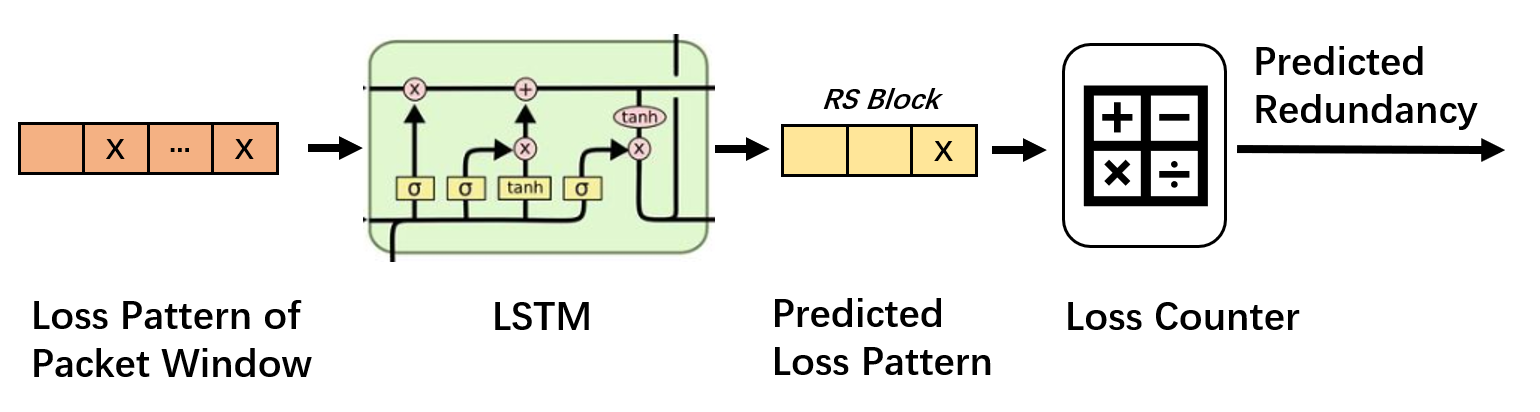}
	
	\caption{The diagram of LSTM Module. LSTM takes the vector of historical pattern as input 
and predicts future pattern, then the Loss Counter accumulates number of lost packets in the future pattern and generate the output.}
	\label{model}
\end{figure}
%The figure of System Architecture. The fundamental information of LSTM.
DeepRS consists of two main modules, LSTM network and RS encoder. 
At the every beginning of encoding procedure, after collecting information from receiver, 
the LSTM module predicts the incoming network packet loss, and then the RS encoder generates 
redundant packets according to the results of LSTM.

\subsection{DeepRS Packet Loss Prediction}
DeepRS takes advantage of LSTM model to predict network packet loss.
LSTM \cite{lstm} is a very popular category of deep learning model, which 
has been proven to have powerful capability to process sequenced
data, for example, the word sequences in natural language.
An LSTM cell mainly contains three gates: forget gate, input gate and output gate.
These gates are responsible for eliminating useless historical information,
updating the current state of cell according to the input and generating output separately.
The particular structure enables LSTM to unearth the contextual relevance
of sequential data more efficiently.

The contextual correlation between historical packet loss pattern and future packet loss
pattern is a natural idea. Historical packet loss pattern implicitly indicates the state of
the network, which plays a leading role in future packet loss. This intuition inspires 
us that applying LSTM model can be a wonderful method to make prediction on network packet loss. 
%In the field of Network, LSTM is confirmed to be an excellent tool to deal
%with learning problems derived from dynamic change of network. 
%What's more, 
Moreover, It has been clarified that network conditions do not change drastically
in a short period \cite{ganjam2015c3} and packets share similar states under 
the same network conditions \cite{jiang2013shedding}, which indicates that 
network fluctuation possibly has regularity to some extent. These conclusions are firm support 
that encourages us to use learning based method, or rather, LSTM model, to predict 
the packet loss of next sending block by learning from the historical loss pattern.

The structure of LSTM module is shown in Fig. \ref{model}. LSTM takes the historical
loss pattern as input and outputs the predicted packet loss pattern. In the training
step, a large amount of historical packet loss sequences are collected and split into
blocks containing $6$ packets. Each sample of data set contains packet loss pattern of 
$5$ blocks as input and the loss pattern of the next block as label. In the reference 
stage, LSTM module predicts the packet loss pattern of incoming coding block based 
on the feedback of $5$ blocks from receiver.
% In view of the above-mentioned facts, we decide to make good use of LSTM to help 
% solving the problem of packet loss in multimedia streaming. 
% Unlike traditional FEC systems which only concentrate on the coding algorithms on the sender-side, 
% DeepRS collects information from receivers and uses the LSTM model to predict incoming network 
% packet loss. Afterwards, the output results will instruct
% the RS Code module to select appropriate parameters for next encoding window.
%[The reason why we use LSTM: CFA \cite{jiang2016cfa}, 
%network conditions do not change a lot in a short period \cite{ganjam2015c3}, 
%packets share similar states under the same network conditions \cite{jiang2013shedding}. ]

\begin{figure}[t]
	%\vspace{0.1in}
	\centering
	\includegraphics[width=0.45\textwidth]{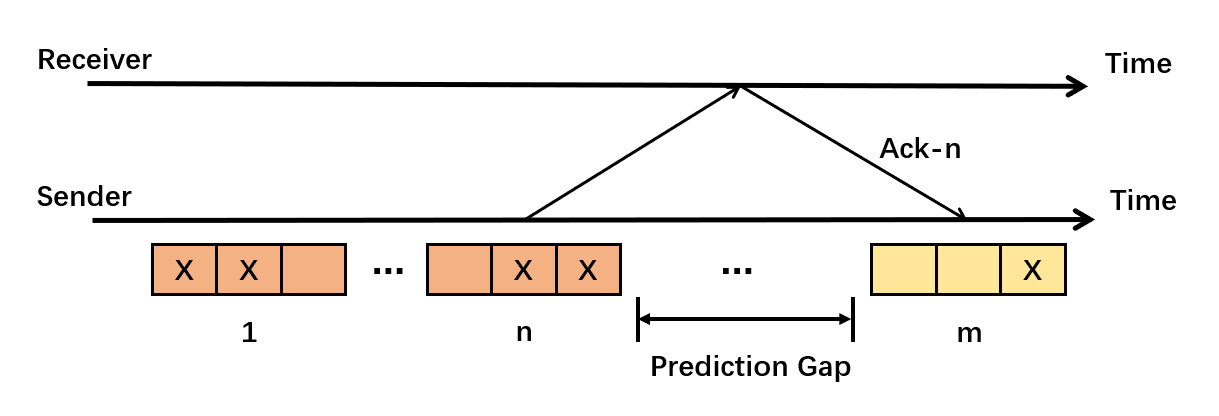}
	%\vspace{-0.85em}
	\caption{Gap resides between the input vector and the output vector.
	This design makes DeepRS adapted for real-time video streaming.}
	\label{io}
\end{figure}

% In our implementation, we choose the packet
% loss pattern of $5$ RS blocks as the input sequence of LSTM.
% The contextual relevance of network conditions 
% is not as strong as natural language in which the deduction is accomplished
% by referring to information several sentences, paragraphs, even pages before. In terms of 
% network conditions, the information of most recent packets matters more usually,
% thus we choose $5$ RS blocks as the input of LSTM, which is suitable to learn hidden features.
% Specifically, blocks related to input sequence are not adjacent to the block related to the output.
% There is another block between input blocks and output block.
% To address the mapping our model learned:
% \begin{equation}
%     f : \{0, 1\}^{5b} \to \{0, 1, 2, ..., b\}
% \end{equation}
% where $b$ indicates the RS block length.

\subsection{Unpredictable Loss Pattern}
General LSTM network is used to learn a mapping from a sample vector to a target vector, 
but the form of output is not suitable for solving the problem of packet loss. 
According to the network conditions, the amount of lost packets in an incoming block 
can be predicted by learning from the historical pattern, but the loss state of each packet 
is harder to be exactly determined because of randomness. What the LSTM model learns from 
is the historical loss pattern, which is mainly influenced by the network conditions, so 
that it can reflect the future packet loss to some extent. Nevertheless, the loss position 
does not have certain relationship with the historical loss pattern. In a word, the future loss 
pattern is sometimes unpredictable, which makes it improper to be the final output.

Fortunately, from the perspective of FEC, actually we do not need to know the loss state for each packet accurately.
This character enables us to avoid the difficulty of directly predicting loss pattern.
We use Loss Counter to predict the number of packets lost in each block based the loss pattern as the final output. 
Since it is directly affected by the network conditions just like the historical loss pattern, 
the count of packet loss can be exactly derived by the fully trained LSTM network.

As is shown in Fig. \ref{model}, we attach Loss Counter after LSTM, transforming the 
loss pattern into number of lost packets. According to the predicted number of lost packets, 
DeepRS decides how many redundant packets ought to generated in this block.
%What we are concerned about is the count of losses in this block, which decides the parameters of FEC packets. 

%As a result, we attach a fully connected layer to the end of LSTM model,
%reducing the dimension of the output to one. After this reforming of LSTM network, 
%the concrete design of this module is shown in Fig. \ref{model}.

% \begin{figure}[t]
% \vspace{0.1in}
% \centering
% \includegraphics[width=0.48\textwidth]{pdf/pred.png}
% \vspace{0.1in}
% \caption{Integral Hierarchy of DeepRS}
% \label{pred}
% \end{figure}

\subsection{Delayed Feedback}
Because of the existence of RTT, the packet loss pattern feedback from receiver is delayed.
DeepRS is not able to obtain the newest network status when a video block is about to be 
sent. Therefore, we need to make prediction on incoming packet loss pattern according to 
packet loss an RTT ago.

To handle this problem, we propose the block gap method for prediction and
inference, which is shown in Fig. \ref{io}. In the training step, we insert a Prediction Gap between
the historical loss pattern and the blocks to be predicted. The Prediction Gap contains some blocks, whose sending
time is an RTT in total, to analog the effect of network delay. In the inference step, DeepRS can still
work in spite of the existence of RTT thanks to the design of Prediction Gap in the training step.

\label{rs}

\begin{figure}[t]
    \centering
    \includegraphics[width=0.36\textwidth]{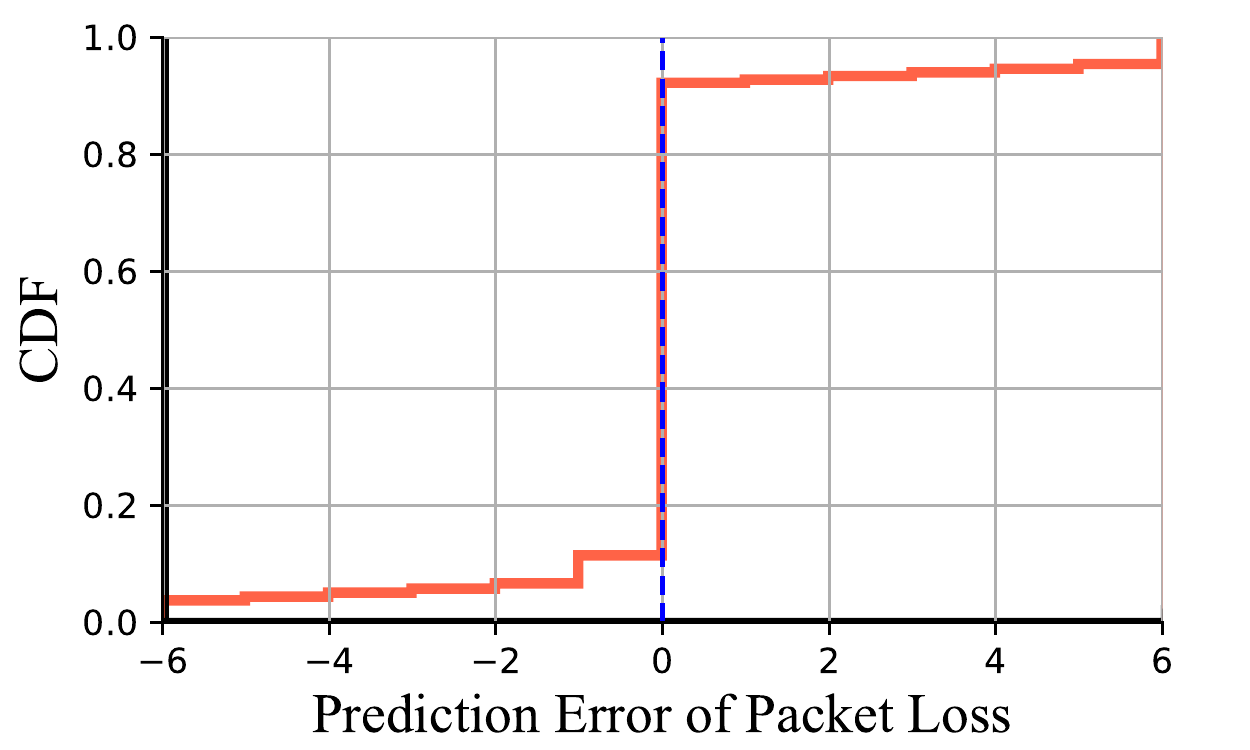}
    \caption{Prediction error distribution of DeepRS under simulation traces indicates 
	the majority of predictions are completely accurate.}
    \label{simcdf}
\end{figure}

%% file: eva.tex
\section{Experiments and Evaluations}
\label{eva}
To validate the efficiency of DeepRS, we carry out trace-based evaluations, 
both on simulation traces and real Internet packet loss traces.

\subsection{Setup}
\subsubsection{Dataset}
Gilbert-Elliot (GE) channel is acknowledged as a common simulation environment
of network packet delivery.
%due to its capability to imitate both random and 
%burst packet loss behaviors of network. 
In Section \ref{sim}, we carry out experiments on simulation
traces generated by GE channel and evaluate the performance of DeepRS.

Real Internet packet loss traces published by 
Fu et.al \cite{Fu2015Experimental} contains transmission meta data of tens of thousands of 
packets in 802.15 WPAN network. The performance of DeepRS on the Internet packet loss dataset 
is evaluated in Section \ref{trace}.

\subsubsection{Underlying FEC algorithm}
As the name of DeepRS indicates, RS code is selected as the underlying FEC algorithm. 
In our implementation, $b$ source packets are grouped as a block and the encoder 
generates $k$ redundant packets based on these source packets.
Once $b$ or more packets, including both source and redundant packets, are collected by 
the receiver, the original data can be recovered by solving a matrix equation. 
For simplicity, we assume redundant packets will not be lost.
% The encoding equation is formulated as followed:
% $$F_{k\times l} = M_{k\times b} \cdot S_{b\times l}$$
% where 
% $$S_{b\times l} = [s_1, s_2, ..., s_{b}]^T$$
% $$F_{k\times l} = [f_1, f_2, ..., f_{k}]^T$$
% and column vectors $s_*$ and $f_*$ represent padded source packets and redundant packets separately.

We choose naive RS methods as the contrast algorithms, since they are widely used in multimedia streaming. 
We use \textbf{Fix-* RS} to represent these algorithms, while the redundancy ratio is static. 
* indicates the fixed redundancy ratio.

\begin{figure}[t]
	\centering
	\includegraphics[width=0.48\textwidth]{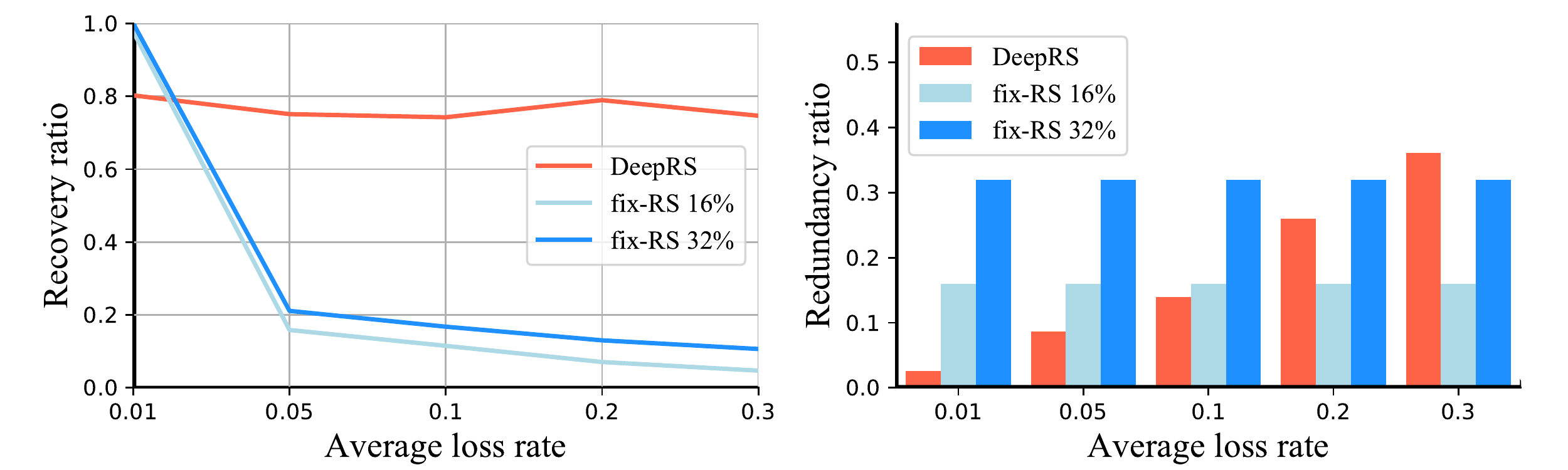}
	
 \ \ \ \ \ \ \ \ \ (a) Recovery ratio\ \ \ \ \ \ \ \ \ \  (b) Redundancy ratio
	\caption{Performance comparison under different loss rate shows that
    DeepRS keeps its performance stable by adjusting its redundancy ratio.}
	\label{simloss}
\end{figure}
\begin{figure}[t]
	\centering
	\includegraphics[width=0.48\textwidth]{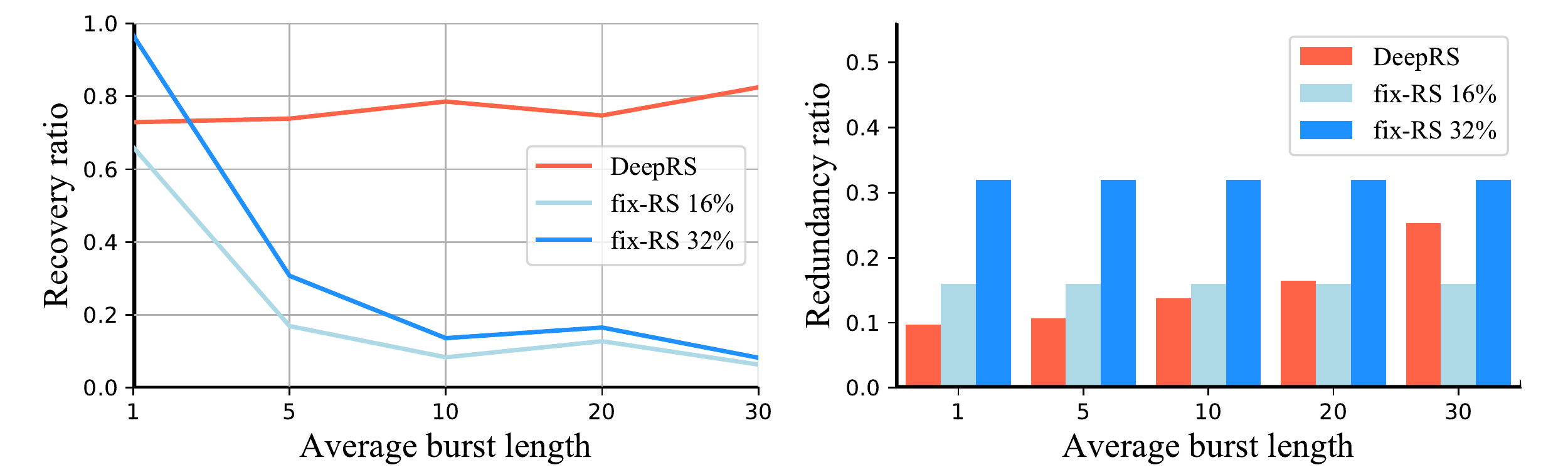}
	
 \ \ \ \ \ \ \ \ \ (a) Recovery ratio\ \ \ \ \ \ \ \ \ \  (b) Redundancy ratio
	\caption{Performance comparison under different burst length also shows 
    the stable performance and the adaptive redundancy ratio of DeepRS.}
	\label{simburst}
\end{figure}
\subsubsection{Metrics}
In performance comparison, we take the following measurement metrics into consideration:
\begin{itemize}
    \item \textbf{Recovery Ratio}. The ratio of recovered packets to all lost packets.
    For instance, recovery ratio is $1$ when all lost packets are recovered.
    Conversely, recovery ratio is $0$ if all lost packets cannot be recovered.
    \item \textbf{Redundancy Ratio}. The ratio of redundant packets to source packets.
    For instance, if RS module generates $k$ FEC packets with a block including $b$ source
    packets, the redundancy ratio is $\frac{k}{b}$.

\end{itemize}

\subsection{Experiments on Simulation}
\label{sim}
Simulation
traces contain 10,000 samples generated from the output of GE channel. 
For each sample, we collect the output vector of length $7b$, then take 
the front part of length $5b$ as input vector, ignore the Prediction Gap whose length is $b$, 
and take the last part of length $b$ as label. 

\subsubsection{Analysis of Prediction Error Distribution}
\label{sec_simcdf}
In this section, experiments are demonstrated on a GE channel given a group of fixed parameters.
%First, We follow the steps described above to generate the data set.
Simulation traces are split into training set, validation set and test set
which account for 60\%, 20\%, 20\% of whole data set separately.
Finally, results on test set are illustrated in Fig. \ref{simcdf} after 
the loss function converges on validation set.
% 需要在前面定义好prediction error

According to the results shown in Fig. \ref{simcdf}, DeepRS predicts the
number of packet loss accurately (zero error) with 70\%$\sim$80\% probability.
% What's more, the prediction error is non-negative with nearly 90\% probability,
% which means DeepRS ensures packet recovery under 90\% circumstances.
This result validates our insight preliminarily.

\begin{figure}[t]
    \centering
    \includegraphics[width=0.36\textwidth]{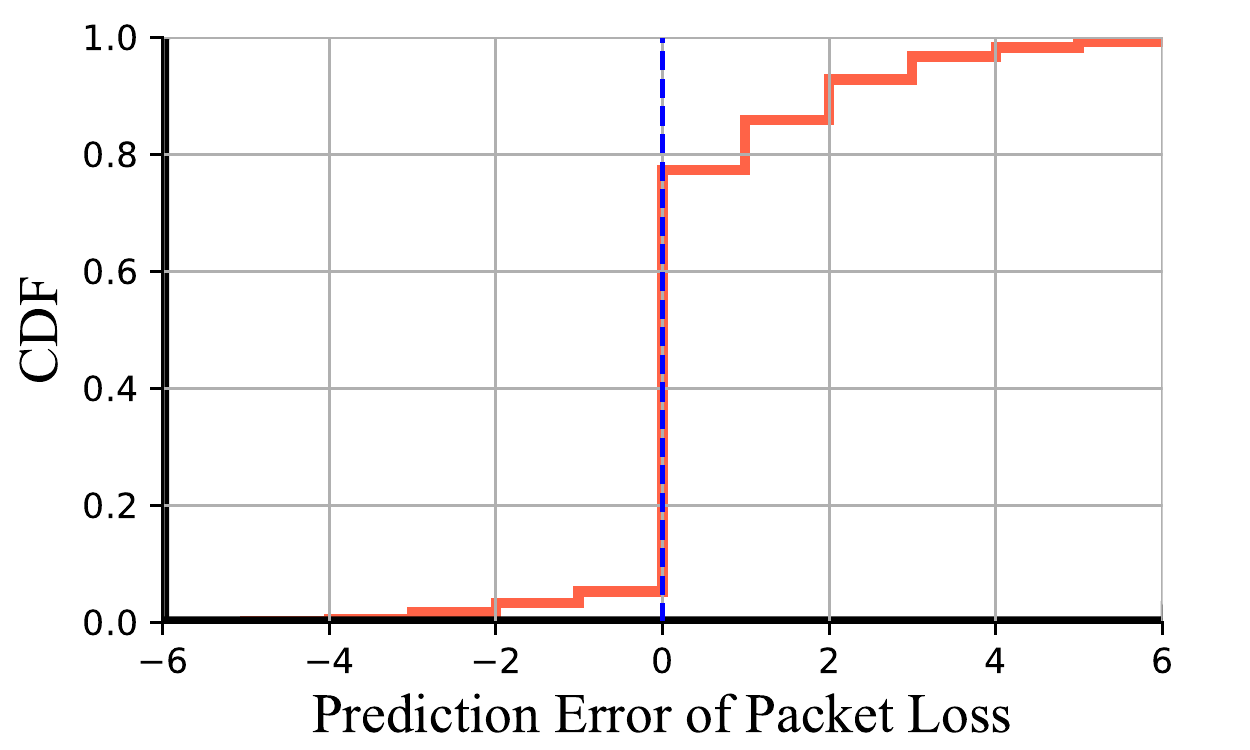}
    \caption{Prediction error distribution under real Internet traces
    also shows high prediction accuracy of DeepRS.}
    \label{sum}
\end{figure}

\subsubsection{Performance under Different Loss Rate}
In this section, we evaluate the performance of DeepRS with diverse average loss rate 
given fixed average burst length.
We choose fix-16\% RS and fix-32\% RS as contrast algorithms. 
%to throw light on the performance gap
%between our learning based algorithm and baselines.
Let average burst length be 10 packets, average loss rate
1\%, 5\%, 10\%, 20\%, 30\% are selected. The trend of recovery ratio
%which means the ratio of recovery packet number to all lost packet number,
and redundancy ratio 
%which means the ratio of fec packet number to 
%source packet number, 
are shown in Fig. \ref{simloss}.

According to the results, DeepRS
is able to alter redundancy ratio under distinct average loss rates.
As a result, its recovery ratio stays relatively stable while 
the performance of fixed redundancy methods plummet as average loss rate grows up. 
% Particularly, when average loss rate is 30\%, fix-RS 32\% method cannot 
% recover 10\% of lost packets, while DeepRS can recover up to 80\%.
In a word, DeepRS outperforms contrast algorithms greatly.
\label{sec_simloss}

\subsubsection{Performance under Different Burst Length}
Similar to Section \ref{sec_simloss}, experiments are made under
several average burst lengths given a fixed average loss rate.
Average loss rate is set to 10\%, and average burst length varies 
from $1$ to $30$, while the same contrast algorithms are selected.
% The y-axes of Fig.\ref{simburst} share same definition with axes
% in section.\ref{sec_simloss}.

The results in Fig.\ref{simburst} show that the behavior 
of DeepRS is similar to Section \ref{sec_simloss}. According to Fig. \ref{simburst}(b), 
it is clear that DeepRS adjusts its redundancy ratio dynamically under
GE channels with different average burst lengths. Fig. \ref{simburst}(a) shows that DeepRS keeps
high recovery ratio while the contrast algorithms work worse
if average burst length is getting longer.
%To conclude this figure, DeepRS performs far better then the contrast algorithms in this section.
\label{sec_simburst}

\subsection{Experiments on Real Internet Traces}
\label{trace}
In this section, we extract packet loss information from the Internet traces. In total, we split 20192 traces 
into training part, validation part and test part, which account for 60\%, 20\% and 20\% 
of original data separately. Next, we apply sliding window approach to generate samples
from each part of data.
% The samples of Internet packet loss traces share consistent format
% with samples in Section \ref{sim}. 
%Samples from each part make up training set, validation set and test set.
The same standard machine learning process is carried out as mentioned in Section \ref{simcdf}.
Prediction error distribution is shown in Fig. \ref{sum}, which 
indicates that DeepRS is prone to predict number of packet loss accurately
with approximately 70\% probability. 
% As for the rate of recovery,
% 90\% is guaranteed. 
This figure implies that DeepRS also works well in real Internet environment.

\begin{figure}[t]
    \centering
    \includegraphics[width=0.365\textwidth]{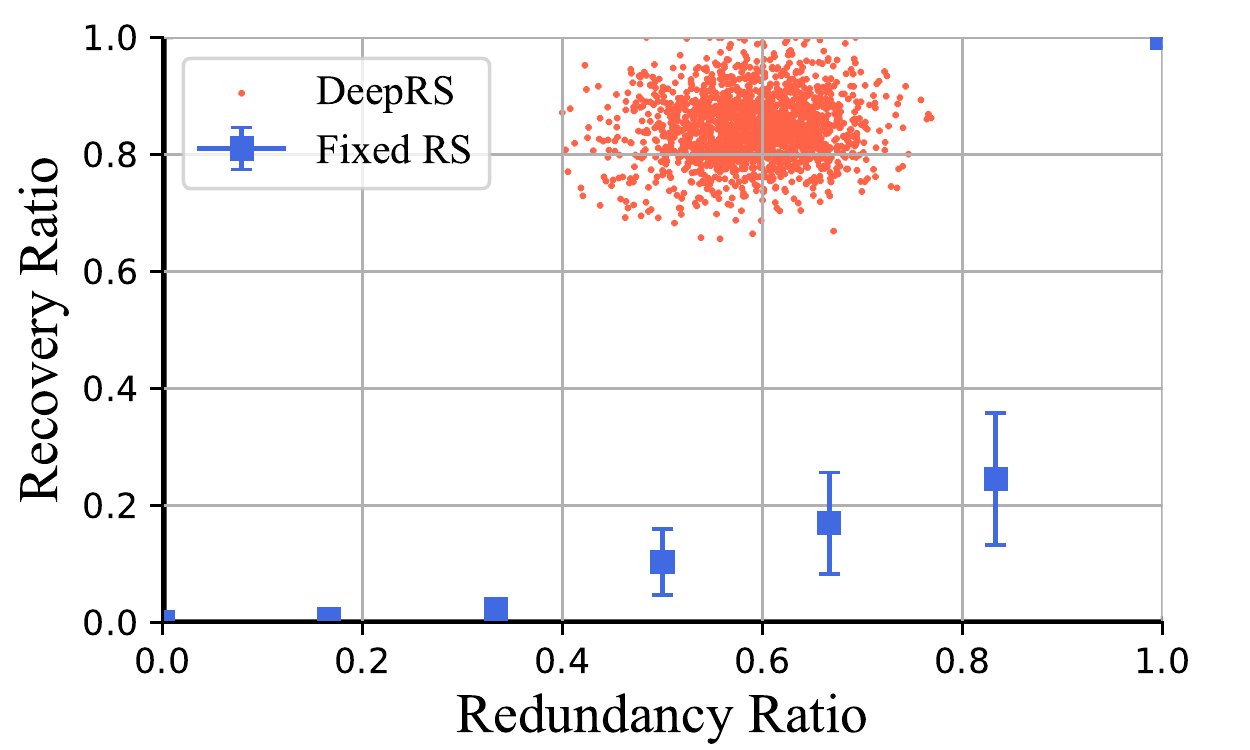}
    \caption{Performance evaluation. DeepRS applies different redundancy ratio for each trace, so the results are shown as 
    discrete points. For each fixed RS method, the 95\% confidence interval is plotted.}
    \label{dot}
\end{figure}

What's more, the trade off between recovery ratio and redundancy ratio
is explored and the performance of DeepRS and fix-RS are 
shown in Fig.\ref{dot}. In this experiment, $7$ distinct fix-RS methods are
selected as contrast algorithms. As is revealed in Fig. \ref{dot}, DeepRS
significantly outperforms traditional RS code methods with fixed redundancy.
DeepRS recovers approximately 80\% packets while fix-RS methods 
help little on packet recovery. 
Notice that the recovery ratio of 
DeepRS is multiple of fix-RS method when redundancy ratio is equal, 
which means that DeepRS is able to achieve higher recovery ratio
with much less extra bandwidth. 
%[95\%]

In conclusion, we make experiments on data set generated by simulation as well as dataset from real world,
and the evaluation of these experiments further proves the superiority of DeepRS.

%% file: conclu.tex
\section{Conclusion}
\label{conclu}
In this paper, we propose DeepRS, a deep learning based FEC system for real-time 
video streaming which mainly consists of two parts, LSTM model and RS encoder. 
DeepRS embeds an LSTM model inside and solves the problem of predicting the number of lost packets in the near future according to the historical packet loss pattern. 
With the help of LSTM model, DeepRS is able to automatically reduce redundancy ratio 
so as to prevent bandwidth waste under low packet loss rate, and increase redundancy 
ratio when packet loss is happening frequently. As for implementation, we have 
modified the LSTM model, thus adapting it to real scenarios in application. 
According to the results of experiments and evaluation, DeepRS achieves far
better trade off between redundancy ratio and recovery ratio than traditional 
fix-redundancy FEC schemes.